\documentclass{emulateapj}

\usepackage{amsmath}
\usepackage{xspace} 
\usepackage{graphicx}
\usepackage{txfonts}
\usepackage{natbib}

\shorttitle{General Relativistic Modulations}

\begin{document}

\title{General Relativistic Flux Modulations from Disk Instabilities
  in Sagittarius A*} 

\author{Maurizio Falanga,\altaffilmark{1,2} Fulvio Melia,\altaffilmark{3,4} 
Michel Tagger,\altaffilmark{1,5} \\ Andrea Goldwurm,\altaffilmark{1,5}
and Guillaume B\'elanger\altaffilmark{6}} 

\altaffiltext{1}{CEA Saclay, DSM/DAPNIA/Service d'Astrophysique, 
91191 Gif-sur-Yvette, France; mfalanga@cea.fr}
\altaffiltext{2}{AIM - Unit\'e Mixte de Recherche CEA - CNRS -
  Universit\'e Paris 7 - UMR n$^{\circ}$ 7158} 
\altaffiltext{3}{Physics Department and Steward Observatory, The
University of Arizona, Tucson, AZ 85721}
\altaffiltext{4}{Sir Thomas Lyle Fellow and Miegunyah Fellow.}
\altaffiltext{5}{Universit\'e Paris Diderot-Paris 7 et Observatoire de Paris,
Laboratoire APC, Paris, France}
\altaffiltext{6}{ESA/ESAC, Apartado 50727, 28080 Madrid, Spain.}

\begin{abstract}

Near-IR and X-ray flares have been detected from the supermassive black 
hole Sgr A* at the center of our Galaxy with a (quasi)-period of $\sim 
17-20$ minutes, suggesting an emission region only a few Schwarzschild 
radii above the event horizon. The latest X-ray flare, detected with 
{\it XMM-Newton}, is notable for its detailed lightcurve, yielding not 
only the highest quality period thus far, but also important structure 
reflecting the geometry of the emitting region. Recent MHD simulations of 
Sgr A*'s disk have demonstrated the growth of a Rossby wave instability, 
that enhances the accretion rate for several hours, possibly accounting 
for the observed flares. In this {\it Letter}, we carry out ray-tracing 
calculations in a Schwarzschild metric to determine as accurately as 
possible the lightcurve produced by general relativistic effects during 
such a disruption. We find that the Rossby wave induced spiral pattern 
in the disk is an excellent fit to the data, implying a disk inclination 
angle of $\approx 77^\circ$. Note, however, that if this association is 
correct, the observed period is not due to the underlying Keplerian motion 
but, rather, to the pattern speed. The favorable comparison between the 
observed and simulated lightcurves provides important additional evidence 
that the flares are produced in Sgr A*'s inner disk.
\end{abstract}

\keywords{accretion---black hole---Galaxy: 
center---instabilities---magnetohydro\-dynamics---relativity}

\section{Introduction}
Long-term monitoring of Sgr A* has thus far resulted in the discovery
of several near-IR \citep{Genzel03} and X-ray flares \citep{Baganoff01} from 
this object. One of the most recent events, observed with {\it XMM-Newton} in 
2004, has produced compelling evidence of a significant modulation in the 
X-ray lightcurve with a quasi-period $\sim 22.2$ minutes \citep{Belanger06}. 
Similar periodic fluctuations had also been seen earlier in the near-IR events.

A period of $\sim 22.2$ minutes is rather intriguing because simple
considerations \citep{MeliaNature01} would place the corresponding emission 
region at roughly 3 Schwarzschild radii ($r_s\equiv 2GM/c^2$) for a black hole 
(BH) mass of $\sim 3.4\times  10^6$ M$_\odot$ \citep{Schodel03}. It is
not yet understood, however, why the Keplerian radius corresponding to
this period is actually less than $3r_s$.  
It could be that the difference is due to the spin of the BH, which 
moves the inner disk radius inwards for prograde rotation. Or it could be that 
the so-called stress edge---the location where the inspiralling material 
actually detaches from the rest of the magnetized disk---is dynamically 
important in establishing where the disk emission terminates
\citep{KrolikHawley02,Melia07}.

Part of the uncertainty is due to the fact that the structure of the disk 
surrounding Sgr A* is itself not fully understood yet. We have recently begun 
to simulate the behavior of the hot, magnetized disk during a disruptive event 
that may be responsible for the flares, under the assumption that the 
instability is induced by low angular momentum clumps of plasma ``raining" 
inwards towards the Keplerian region (Tagger \& Melia 2006; see also Chan et
al. 2006).  This scenario is motivated 
by extensive hydrodynamic \citep{MC99,CN05} and MHD \citep{IN02} simulations 
which show that for the stellar-wind fed conditions at the Galactic center, 
the average specific angular momentum of gas captured gravitationally by
Sgr A* is too small to sustain a  `conventional' (i.e., typically large 
$\sim 10^5\,r_s$) disk. Instead, only clumps of plasma with relatively small 
angular momentum venture inwards and merge with---essentially, `rain'
onto---the compact disk at the circularization radius, which for Sgr A* is $\la
10-10^3\,r_s$ \citep[see also][]{MF01,M07}. 

These MHD simulations \citep{TM06} did indeed establish the result
that the merger of an infalling blob with the existing Keplerian flow induces 
a Rossby wave instability (RWI) leading to the total disruption of the disk on a 
timescale relevant to the Sgr A* flare phenomenon. But unless the ensuing 
X-ray modulation is due to a QPO response in the inspiralling matter (see
Chan et al. 2006), one 
cannot escape the fact that general-relativistic effects are essential in 
producing a periodicity in the lightcurve due to radiation by an azimuthally 
asymmetric emitter. The pattern of modulation over one complete cycle would be 
the result of several influences, including a Doppler shift, light-bending, 
and lensing effect near the BH's event horizon \citep[e.g.,][]{HM95}. 

In this {\it Letter}, we take the results from the (non-relativistic)
MHD simulation of \citet{TM06} and carry out a full ray-tracing calculation 
of the lightcurve produced by the disruption for direct comparison with the 
{\it XMM-Newton} data. 

\section{Observed Lightcurve of an X-ray flare from Sgr A*}
An inspection of Fig. 5 in \citet{Belanger06} shows that the effects
of gravitational light-bending, lensing, Doppler effect and travel
time delay may have helped to shape the folded  
lightcurves from the 2004 August 31 event. One cannot yet discount the
possibility that an actual periodic dynamical effect may have also contributed 
to the modulation seen in the emissivity, but this is necessarily 
model-dependent and the signature may not be unique. For example,
Tagger \& Melia (2006) found that a quasi-periodic modulation could
be excited by non-linearities in the evolution of the spiral-Rossby
pattern, which would sit on top of the modulation one would see due
to general relativistic effects. For the purposes of this 
{\it Letter}, we will adopt the simplest assumption---that the
modulation is due predominantly  to general relativistic effects. 

The 2--10 keV folded curve from the 2004 August 31 event (see 
Fig. \ref{fig:fig3}) looks broader and, except for one significant datum  
standing well away from the rest, looks more like a continuous, relatively 
smooth modulation rather than the sharp changing profile produced by an 
orbiting hot spot \citep[see e.g.,][]{HM95}. It is partially for this 
(phenomenological) reason that the Rossby wave instability is 
promising and worth investigating further here, because the disruption it 
causes is global, offering the possibility of producing a gradual modulation 
in the lightcurve when general relativistic effects are included.

\section{Numerical Simulation of the Instability}
The instability we have simulated with our MHD code has a long history, 
dating back to \citet{LH78}, who showed that a disk presenting an extremum of 
a quantity $\cal L$ (later dubbed vortensity) was subject to a local
instability of Rossby vortices. The requirement of an extremum is similar to 
that giving rise to the Kelvin-Helmholtz instability of sheared flows. More 
recently, \citet{Lovelace99} renamed it the Rossby Wave Instability (RWI) 
and developed the theory and numerical simulation.  

In isothermal, unmagnetized disks, $\cal L$ is the specific vorticity
averaged across the disk thickness,
\begin{equation}
{\cal L}\ =\ \frac{\vec\nabla \times \vec V}{\Sigma}\ =\
\frac{\kappa^2}{2\Omega\Sigma}\;, 
\end{equation}
where $\Sigma$ is the disk's surface density, $\Omega$ its rotation
frequency, and
\(
\kappa^2\ =\ 4\Omega^2+2\Omega\Omega'r
\)
is the epicyclic frequency squared.
The extremum of $\cal L$ appears to be due to an extremum in the radial density
profile. To understand how the instability is driven, we note that Rossby 
waves in disks form normal modes trapped near the extremum of $\cal
L$. In the MHD form of the RWI, the disk is threaded by a vertical
(poloidal) magnetic field $B_{0}(r)$. Its properties are essentially
the same as those discussed above, except that here the critical quantity is 
\(
{\cal L}_{B}={\kappa^2\Sigma}/({2\Omega}{B_{0}^2}),
\)
and the growth rate can be higher because of the long-range action of
the Lorentz force coupling the Rossby vortices. 

\section{Ray Tracing Calculations}
\label{GR-RT}
A typical profile of the inner disk during the Rossby wave growth is shown
in Fig. 3 of \citet{TM06}. In this paper, we present the 
lightcurve and images associated with this disruption, from the vantage point 
of an observer at infinity. The calculation is carried out with a fully
general relativistic ray-tracing code. The RWI arises in the compact
accretion disk surrounding the (Schwarzschild) BH, and we
describe its morphology using coordinates in the co-rotating frame 
($r,\theta,\varphi$). The modeled accretion disk is thin and the 
RWI may be considered to lie in the equatorial plane ($\theta = \pi/2$)
of the compact object. The observer is located at infinity with
viewing angle {\it i} relative to the $z'$-axis in the non-rotating
frame, at (observer) polar coordinates ($r',\theta',\varphi'$). The deflection angle 
of a photon emitted by plasma in the Rossby-unstable region is $\psi$, 
and varies periodically with $\cos\, \psi = \cos\,i\, \cos \,\varphi$.

These emitted photons are deflected by the BH and intersect the 
observer's detector plane at infinity. The distance between the line-of-sight 
and the point at which the photon reaches the detector is defined as the 
impact parameter {\it b}. Using this geometry, the deflection angle of
the photon's trajectory may be obtained with the light-bending relation
between $\alpha$ (the angle between the emission direction of the photon
and the direction from the center of the BH to the location of
the emitter) and $\psi$ from the geodesic equation
\begin{equation}
\psi = \int_{R}^{\infty} \frac{dr}{r^{2}}\biggr[
  \frac{1}{b^{2}}-\frac{1}{r^{2}}\biggr(1-\frac{r_{s}}{r}\biggl)\biggl]\;.  
\label{eq:psi}
\end{equation}
This procedure yields the impact parameter $b = r\,(1-r_{s}/r)^{-1}
\,\sin\,\alpha$ of the photons in terms of the emitting radius $r$, 
and ultimately allows us to calculate the flux at infinity. 
A detailed description of this geometry is provided in, e.g., 
\citet{Luminet79}, and \citet{Falanga07}. In our derivation, we
shall use the same notation and geometry described in  \citet{pg03}, 
though pertaining to a localized region (or clump) in the disk.
The system of units is chosen such that $G=c=1$; in this standard
coordinate system, the BH's horizon occurs at the 
Schwarzschild radius $r_{s} = 2M$, where $M$ is the mass of the 
compact object.


In the simulations we report here, the emitting region is geometrically
thin, and generally optically thin as well. Our rays include an integration
through the whole emitting depth, but because most of the emissivity is 
concentrated near the disk's plane, we effectively have a situation in 
which the rays themselves appear to begin very close to this plane. Rays 
leaving the disk in directions that eventually take them around the
BH, heading toward the observer, contribute much less to the overall
flux and we ignore them here.   

The general-relativistic effects to be considered are now: ({\it i})
light-bending (see above), ({\it ii}) gravitational Doppler effect
defined as (1+z), ({\it iii}) gravitational lensing effect, 
$d\Omega_{\rm obs}=b\,db\,d\varphi/D^2$ (with $D$ the distance to the source), 
expressed through the impact parameter, and ({\it iv}) the travel time delay.
We calculate the 
relative time delay between photons arriving at the observer 
from different parts of the disk, using the geodesic equation. 
The first photon reaching the observer is the photon
emitted at phase $\varphi=0$, on the closest radius ($r=r_{max}$) 
to the observer. We set this reference time, $T_{0}$, equal to zero. 

\begin{figure*}[ht]
    \epsscale{1.0}
    \begin{center}
       \includegraphics[scale=0.32]{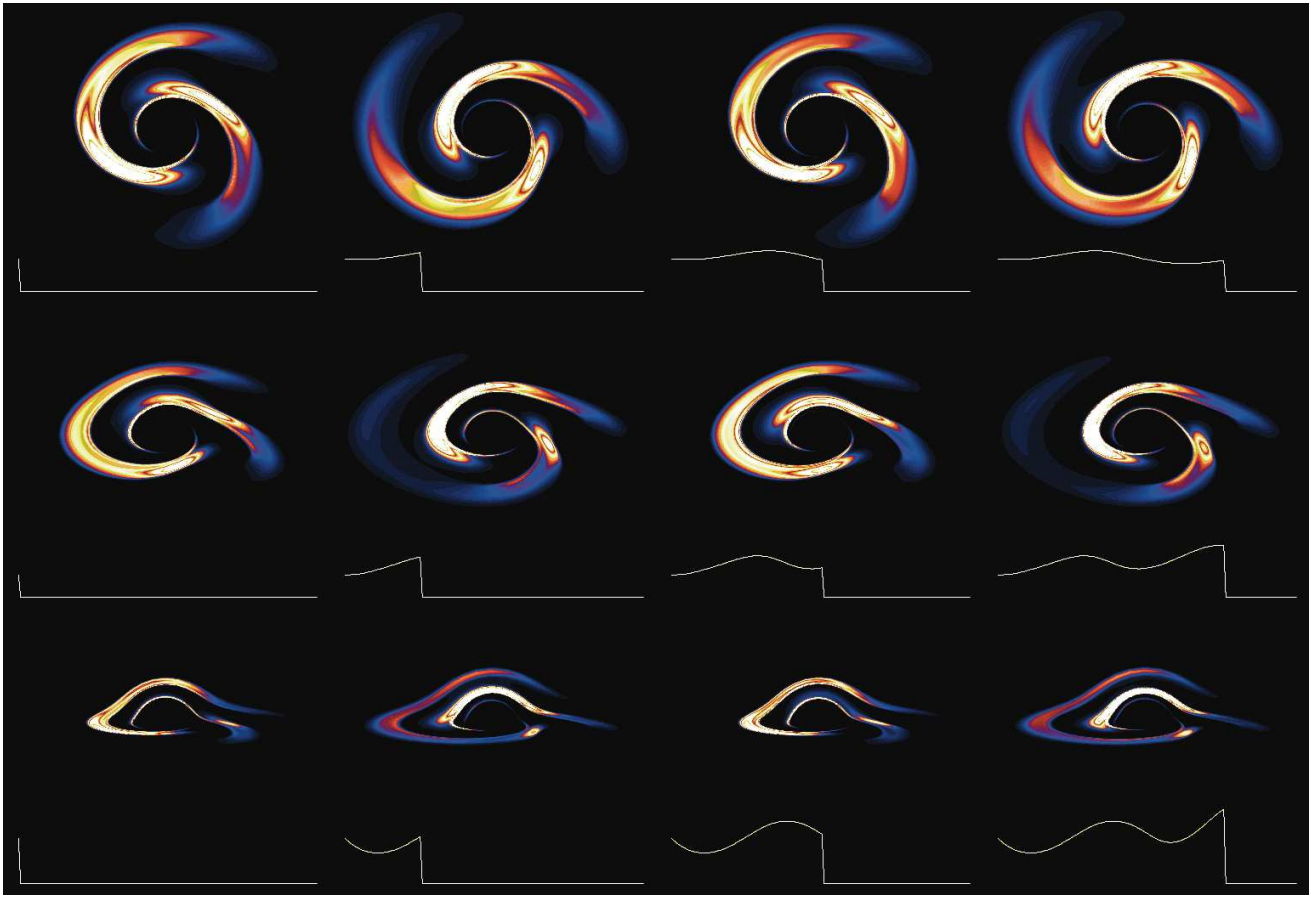}
    \end{center}\vspace{-3mm}
        \caption{\scriptsize Horizontal: Four snapshots of the disk (spiral pattern) as it
	  would appear to an observer looking along a line-of-sight inclined
	  by $30^{\circ}$ (top), $60^{\circ}$ (middle), and $80^{\circ}$ (bottom), relative to
	  the disk's symmetry axis, at phases $\varphi=0,\, \pi/2,\, \pi$, and $3\pi/2$,
	  respectively. Variation in each column is due to general
	  relativistic light-bending, lensing, and Doppler shifts, including
	  the relativistic time delay for the various inclinations. The
	  lightcurves attached to their respective images show how an observer
	  views these overall effects from infinity (see also Fig. 2). Slight
	  differences between the two arms, resulting from the MHD simulations, result in two slightly
	  different peaks in the lightcurve during one full rotation of the pattern. 
}
\label{fig:fig1}
\end{figure*}

The observed flux at energy $E'$ is $F_{\rm obs}(E') =
I_{\rm obs}(E')d\Omega_{\rm obs}$, where $I_{\rm obs}(E')$ 
is the radiation intensity observed at infinity and
$d\Omega_{\rm obs}$ is the solid angle on the observer's sky 
including relativistic effects. Using the relation 
$I_{\rm obs}(E',\alpha')=(1+z)^{-3} I_{\rm em}(E,\alpha)$, 
a Lorentz invariant quantity that is constant along null geodesics in
vacuum, the intensity of a light source integrated over its effective 
energy range is proportional to the fourth power of the redshift
factor, $I_{\rm obs}(\alpha')= (1+z)^{-4}I_{\rm em}(r,\varphi)$, 
$I_{\rm em}(r,\varphi)$ being the intensity measured in the rest 
frame of the clump \citep{MTW73}. 

Although our magnetohydrodynamic simulation, and corresponding ray-tracing
calculation, are quite sophisticated, they are nonetheless still somewhat
restricted in that we have not allowed for a completely self-consistent 
coupling between the plasma and the radiation. The infalling plasma 
radiates inefficiently, so this is not a serious deficiency as far as 
the dynamics is concerned. Furthermore, properly modelling the compression 
of the gas would require fully 3-D simulations, which MHD codes are still 
unable to handle in the conditions (a disk threaded by a near-equipartition 
vertical field) that we use. For simplicity, the MHD simulation was thus carried
out assuming isothermal conditions. However the electron temperature, which
dominates the radiative emission, responds to the compression of the gas
and we model this here with an ad-hoc prescription, from the compression 
obtained in the simulation. This is sufficient to provide the main result 
of the present work, which is the form of the light curve that the spiral 
pattern and general relativistic effects generate. Also, we have not
included non-local effects arising, e.g., from inverse Compton
scattering. In determining the surface emissivity, it therefore makes
sense to take a simplified approach in which we include principally
the parameter scalings, rather than their absolute values. This
procedure will give us correct amplitudes in the lightcurve, though
not the absolute value of the flux per se. 

Using the perfect gas law for an {\it adiabatic} flow (since the gas
is radiatively inefficient), we can invoke a polytropic equation of
state with $\gamma=5/3$ and write the temperature as $T \propto 
\rho^{2/3}$. This assumes further that the radiation pressure is 
negligible and $T$ is not so high that the particles are strongly 
relativistic. Although the disk is not in full hydrostatic equilibrium, 
we can still argue that on average, we should have for a steady thin
disk the gas density $\rho = \Sigma/H(r,z)$, where $\Sigma$ is the
column density, and $H$ is the disk height. These give
$\rho\propto\Sigma^{3/4} r^{-9/8}$ (using $r$ to denote the radius
in the equatorial plane) and $T \propto \Sigma^{1/2} r^{-3/4}$. 
The synchrotron emissivity is therefore 
$j_s \propto B n_{\rm nt} \propto \Sigma (T\rho)\propto 
\Sigma \rho^{(5/3)}$, where the nonthermal particle energy is
roughly in equipartition with the thermal. We argue that the
plasma is fully ionized and resistivity is minimal, so $B$ is frozen
into the gas, which means that $B\propto \Sigma$. We therefore
infer that $j_s \propto \Sigma^{9/4} r^{-15/8}$. 

Now the X-rays are produced via inverse Compton scattering from the  seed
photon number flux. Thus, with $L_{\rm seed}\propto r^3\,  j_s$, where $j_s$
is the synchrotron emissivity in units of energy per unit volume per unit
time, the soft photon flux scales as the emitted power divided by the
characteristic area. That is, $F_{\rm seed}\propto r^3\, j_s / r^2 = r j_s$,
which is going to be roughly the same scaling as the seed photon density, so
$n_{\rm seed} \propto r j_s \propto \Sigma^{9/4} r^{-7/8}$.  The inverse
Compton scattering emissivity is therefore $j_{ic}  \propto n_{\rm nt}\,
n_{\rm seed} \propto \Sigma^{7/2} r^{-11/4}$.  Thus, $j_{X-ray}\sim j_{ic}$,
and the surface intensity is $I_{\rm em} \propto \int j_{X-ray} ds \propto
j_{X-ray} H$, which gives finally $I_{\rm em} \propto \Sigma^{15/4}
r^{-13/8}$.  The core physics in this expression is the surface density
$\Sigma(r,\varphi)$, which we take from the MHD simulation in \citet{TM06}.

In order to evaluate the flux at a given azimuthal angle 
$\varphi$ and radius $r$, we first compute numerically 
$\psi(\alpha)$, and then calculate the Dopper shift, 
lensing effects, and finally the flux $F_{\rm obs}$ as a
function of the arrival time, including all the travel 
time delays described above. The simulation is carried 
out on a polar grid with $n_{r}=256$ and $n_{\varphi}=128$ 
points, extending from $r_{\rm in}=3r_s$ to $90r_s$. In this pass through
the problem, in order to clearly separate the general relativistic 
modulation from the complex dynamical behaviour observed during the 
simulation, we have chosen to take a representative snapshot and rotate 
it at the frequency measured in the simulation. A full treatment of the 
whole simulation will be presented in \citet{Falanga07}. 
\begin{figure}[ht]
    \epsscale{1.0}
    \begin{center}
        \includegraphics[scale=0.4, angle=-90]{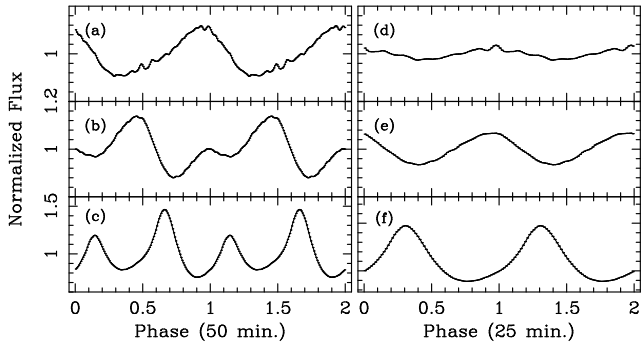}
    \end{center}\vspace{-3mm}
        \caption{\scriptsize Panels (a-c) show the lightcurve observed at infinity,
resulting from general relativistic flux modulations
associated with a Rossby-wave-disrupted disk. The period
arises from the two-spiral-armed pattern speed, and is roughly
50 minutes for this particular MHD simulation. Panel (a) is
for a disk inclination of $30^{\circ}$, (b) is $60^{\circ}$, and (c) is
$80^{\circ}$, in a one-to-one correspondence with the rows of
Fig. 1. Panels (d-f) show the same lightcurve, though now
folded on half of a pattern-rotation period, i.e., 25 minutes. Each period is
repeted once for clarity.
}
\label{fig:fig2}
\end{figure}

\section{Results and Discussion}

Our principal results are presented graphically below, and may be
understood with reference to the simulated images shown in Fig. \ref{fig:fig1}.
These are intensity maps projected onto the plane of the sky,
for the ray-traced perturbed disk in \citet{TM06},
at three different inclination angles: $30^\circ$ in the top row,
$60^\circ$ in the middle, and $80^\circ$ for the bottom row. The
four columns are snapshots taken at 4 (equally-spaced) phases of 
one complete pattern revolution. The general relativistic 
distortions depend strongly on inclination angle, which we 
employ in our search for the best fit to the modulation in the 
X-ray lightcurve.

This variation is demonstrated quantitatively in Fig. \ref{fig:fig2}, which
shows the lightcurves corresponding to the three inclinations
illustrated in Fig. \ref{fig:fig1}. Note that at small inclination, we detect
a gradual, broad modulation, whereas for the higher inclinations,
we begin to see the effects of a two-spiral arm emitting region.
Interestingly, the period associated with the pattern rotation
in the \citet{TM06} simulation is about 50 minutes.
Though this calculation was not optimized to fit the observed
period, the fact that we see a bimodal modulation from the two 
spiral arms suggests that one cycle in the data may in fact
correspond to half a revolution of the pattern. In this figure,
we therefore also show the calculated lightcurve folded over
half a pattern period, corresponding to about 25 minutes.

\begin{figure}[ht]
    \epsscale{1.0}
    \begin{center}
        \includegraphics[scale=0.5]{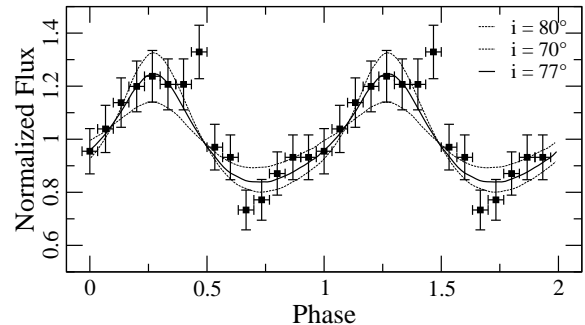}
    \end{center}\vspace{-3mm}
        \caption{\scriptsize Lightcurve of the 2004, August 31 flare
        folded with a phase of 1330~s, in the 2--10~keV energy band
        from \citet{Belanger06}. The best fit model 
is shown by the solid line using an inclination
        angle of $77^{\circ}$. The dasehed line represent upper and
        lower limits using $80^{\circ}$ (bigger amplitude) and
        $70^{\circ}$  (smaller amplitude) inclination
        angles, respectively.}
\label{fig:fig3}
\end{figure}

The true test of relevance for this simulation lies in a direct 
comparison between the data and the inclination-dependent lightcurves. 
We emphasized in the introduction that the various relativistic effects
produce a unique profile, not easily confused with other
periodic modulations, most of which tend to be sinusoidal.
As we demonstrate in Fig. \ref{fig:fig3}, the shape of the calculated
lightcurve, particularly its amplitude, is quite sensitive
to the inclination angle, which again, is most easily understood
with reference to Fig. \ref{fig:fig1}. For example, this figure includes
3 curves, corresponding to inclinations of $70^\circ$,
$77^\circ$, and $80^\circ$. The middle curve produces the best
fit ($\chi^{2}_{red}=1.1$), and it should be noted that the
correspondence to the data is excellent, not only in terms of the modulation 
amplitude, but also for the shape of the lightcurve itself. 
Note that the simulated lightcurve corresponds to the bolometric
flux emission. We have therefore compared the theoretical curve 
with the observed 2--10 keV lightcurve, rather than individually 
in different energy bands, as presented in B\'elanger et al. (2006).

There are several new ideas that we can take away from this work.
First, if the periodic modulation seen in some flares from Sgr A*
by both IR and X-ray instruments is real, it is not at all obvious
that the periodicity is due to an underlying Keplerian period. One
must be very careful, therefore, in over-interpreting these periods
in terms of a BH spin. Second, the shape of the X-ray lightcurve
in this particular flare is too broad for it to correspond easily
to a highly localized hot spot on the disk, which instead would
produce a more strongly peaked profile like that investigated earlier
by \citet{HM95}. Instead, our work here argues for a more global 
disruption in the disk, at least for some events, like the 2004, 
August 31 flare observed with {\it XMM-Newton}.
In this regard, the driving mechanism is likely to be an infall of 
clumps of plasma that merge with the existing compact disk and 
induce a Rossby-type of instability. Finally, the observed power
density spectrum shows that the (quasi)-period is quite clean,
without any evidence that the emission region is spread over a
large range in Keplerian periods. This has been taken as some
evidence that the disturbance must therefore be highly localized,
probably near the marginally stable orbit. But as we have shown here, 
the disturbance need not be that close to the event horizon in order
to produce a modulation with a period of only $\sim 20-25$ minutes.
One may still get a narrow peak in the power density spectrum,
as long as the modulation is due to a pattern rotation, rather than
to motion along an orbit. But this will only work as long as the 
pattern has multiple components, such as the two spiral arms we
have modeled in this paper.

\acknowledgments
The research was partially supported by the French Space Agency (CNES) and
NSF grant AST-0402502 at the University of Arizona. FM is grateful for the
hospitality of the APC in Paris, where most of this work was carried out.


\begin{thebibliography}{}

\bibitem[\protect\citeauthoryear{Baganoff et al.}{2001}]{Baganoff01} 
Baganoff, F. et al. 2001, Nature, 413, 45

\bibitem[\protect\citeauthoryear{B\'elanger et al.}{2006}]{Belanger06} 
B\'elanger, G. et al. 2006. ApJ (Letters), submitted (astro-ph/0604337)


\bibitem[\protect\citeauthoryear{Chan et al.}{2006}]{Chan06}
Chan, C., Liu, S., Fryer, C. L., et al. 2006, \apj, submitted (astro-ph/0611269

\bibitem[\protect\citeauthoryear{Cuadra et al.}{2005}]{CN05}
Cuadra, J., Nayakshin, S., Springel, V. \& Di Matteo, T. 2005, \mnras, 360, L55

\bibitem[\protect\citeauthoryear{Falanga et al.}{2007}]{Falanga07} 
Falanga et al., ApJ, 2007, in preparation

\bibitem[\protect\citeauthoryear{Genzel et al.}{2003}]{Genzel03}
Genzel, R. et al. 2003, Nature, 425, 934

\bibitem[\protect\citeauthoryear{Hollywood et al.}{1995}]{HM95}
Hollywood, J. M., Melia, F., Close, L. M., et al. 1995, \apjl, 448, L21

\bibitem[\protect\citeauthoryear{Igumenshchev \& Narayan}{2002}]{IN02}
Igumenshchev, I.V. \& Narayan, R. 2002, \apj, 566, 137

\bibitem[\protect\citeauthoryear{Krolik \& Hawley}{2002}]{KrolikHawley02}
Krolik, J. H. \& Hawley, J. F. 2002, \apj, 573, 754

\bibitem[\protect\citeauthoryear{Lovelace \& Hohlfeld}{1978}]{LH78}
Lovelace, R. V. E., \& Hohlfeld, R. G. 1978, \apj, 221, 51

\bibitem[\protect\citeauthoryear{Lovelace et al.}{1999}]{Lovelace99}
Lovelace, R. V. E., Li, H., Colgate, S. A., Nelson, A. F. 1999, \apj, 513, 805

\bibitem[\protect\citeauthoryear{Luminet}{1979}]{Luminet79}
Luminet, J., -P. 1979, A\&A, 75, 228

\bibitem[\protect\citeauthoryear{Melia}{2001}]{MeliaNature01} 
Melia, F.  2001, Nature, 413, 25

\bibitem[\protect\citeauthoryear{Melia}{2007}]{M07}
Melia, F. 2007, in {\it The Galactic Supermassive Black Hole} (PUP: New York)

\bibitem[\protect\citeauthoryear{Melia \& Coker}{1999}]{MC99} 
Melia, F. \& Coker, R.F. 1999, \apj, 511, 750

\bibitem[\protect\citeauthoryear{Melia \& Falcke}{2001}]{MF01}
Melia, F. \& Falcke, H. 2001, ARAA, 39, 309

\bibitem[\protect\citeauthoryear{Melia et al.}{2007}]{Melia07} 
Melia, F., Prescher, M., B\'elanger, G. \& Goldwurm, A. 2007, ApJL, submitted 

\bibitem[\protect\citeauthoryear{Misner, Thorne, \& Wheeler}{1973}]{MTW73}
Misner, C. W., Thorne, K., S., \& Wheeler, J., A. 1973, {\it Gravitation}
(San Francisco: Freeman)

\bibitem[\protect\citeauthoryear{Poutanen \& Gierli\'nski}{2003}]{pg03}  
Poutanen, J., \& Gierli\'nski, M. 2003, MNRAS, 343, 1301  

\bibitem[\protect\citeauthoryear{Sch\"odel et al.}{2003}]{Schodel03} 
Sch\"odel, R., Ott, R., Genzel, R., Eckart, A., Mouawad, N., \&
Alexander, T. 2003, ApJ, 596, 1015 

\bibitem[\protect\citeauthoryear{Tagger \& Melia}{2006}]{TM06} 
Tagger, M.\& Melia, F. 2006, ApJ, 636, L33

\end{thebibliography}
\end{document}